\documentclass[a4paper]{jpconf}
\usepackage{graphicx}
\RequirePackage{xspace}
\newcommand{\gev}{\ensuremath{\mathrm{\,Ge\kern -0.1em V}}\xspace}
\newcommand{\gevc}{\ensuremath{{\mathrm{\,Ge\kern -0.1em V\!/}c}}\xspace}
\begin{document}
\title{Particle Identification at Belle II}

\author{S Sandilya \\ {\small On behalf of the Belle II Collaboration}}

\address{University of Cincinnati, Cincinnati, Ohio 45221}

\ead{saurabhsandilya@gmail.com}

\begin{abstract}
We report on the charged particle identification (PID) systems for the upcoming Belle II experiment. 
The time of propagation counter in the central region and the proximity focusing ring imaging Cherenkov counters with aerogel radiator in the forward region will be used as the PID devices. 
They are expected to provide a kaon identification efficiency of more than 94\% at a low pion misidentification probability of 4\%. 
The motivation for the upgrade, method and status of both systems are discussed. 
\end{abstract}

\section{Introduction}
\label{intro}
Reliable particle identification (PID) is essential for any high energy physics experiment, and in particular, it is crucial for the success of $B$-factories~\cite{krizan:2009}.
In the $B$-factories, PID is required to tag $B$-meson flavour for CP violation studies in the neutral $B$-meson system, and to suppress backgrounds in precision measurements of rare $B$ and $D$ decays.

\begin{figure}[h]
\centering
\includegraphics[width=15cm]{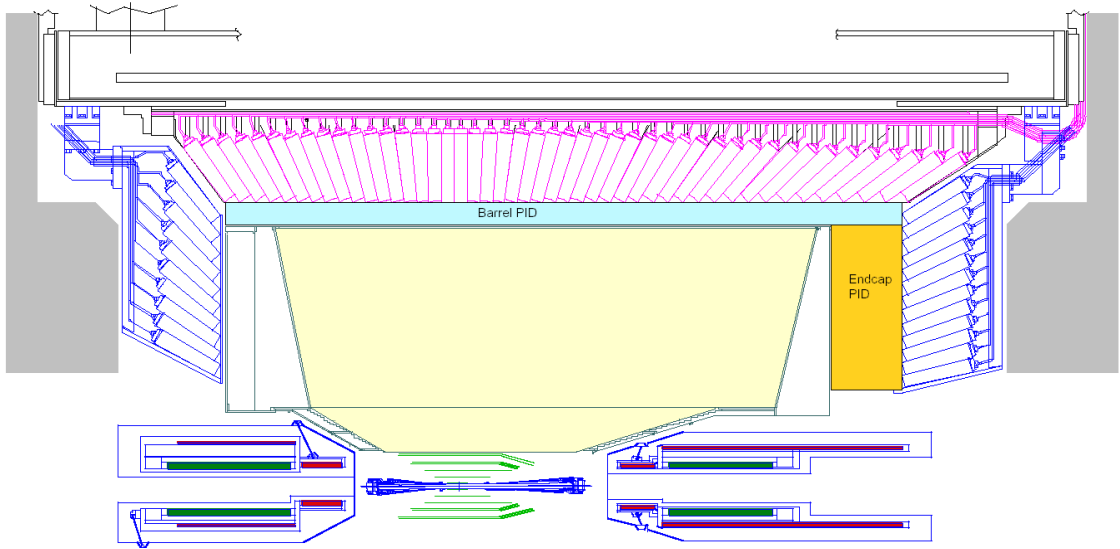}
\caption{Schematic of the upper half vertical cross-section of the Belle II detector. The rectangular region shaded with cyan color represents the TOP counters in the barrel region and the orange shaded rectangular region represents ARICH counters in the forward endcap region.}
\label{fig:pid}
\end{figure}

Based on the successful operation of the Belle experiment, the KEKB~\cite{kekb} is being upgraded to the SuperKEKB collider, which is designed to give an instantaneous luminosity of $8 \times 10^{35} {~\rm cm^{-2}s^{-1}}$. 
This is about 40 times larger than its predecessor. 
The Belle detector~\cite{belle} is also being upgraded to the Belle II detector~\cite{belle2:2011} in order to be able to perform in the upgraded accelerator conditions and to achieve the desired physics goals. 
To cope with high rates as well as with more stringent requirements on separation capabilities for rare decay channels, more efficient PID is needed for the momentum range up to 4~\gevc. 
In Belle II, PID will be performed by the Time-Of-Propagation (TOP) counter in the central (barrel) region and the Aerogel Ring Imaging Cherenkov (ARICH) counter in the forward endcap region. 
Figure~\ref{fig:pid} shows a schematic of the upper half vertical cross-section of the Belle II detector in which both the PID systems are shown.

The working principle of both the TOP and ARICH counters is based on imaging the Cherenkov rings. 
The Cherenkov photons are emitted by a charged particle moving with a velocity $(v = \beta c)$ exceeding the speed of light $(c/n)$ in a dielectric medium (refractive index $n(\lambda)$) at a characteristic angle ($\theta_{c}$) given by, $ \cos\theta_{c} = 1/n(\lambda)\beta $. 
Charged particles of the same momentum but having different masses will have different velocities, and hence the angle of Cherenkov photons will be different.

In this proceedings, we report the design, method and present status of the PID systems in both the regions of the Belle II detector.

\section{TOP counters}
\label{top}
\begin{figure}[h]
\centering
\includegraphics[width=8cm]{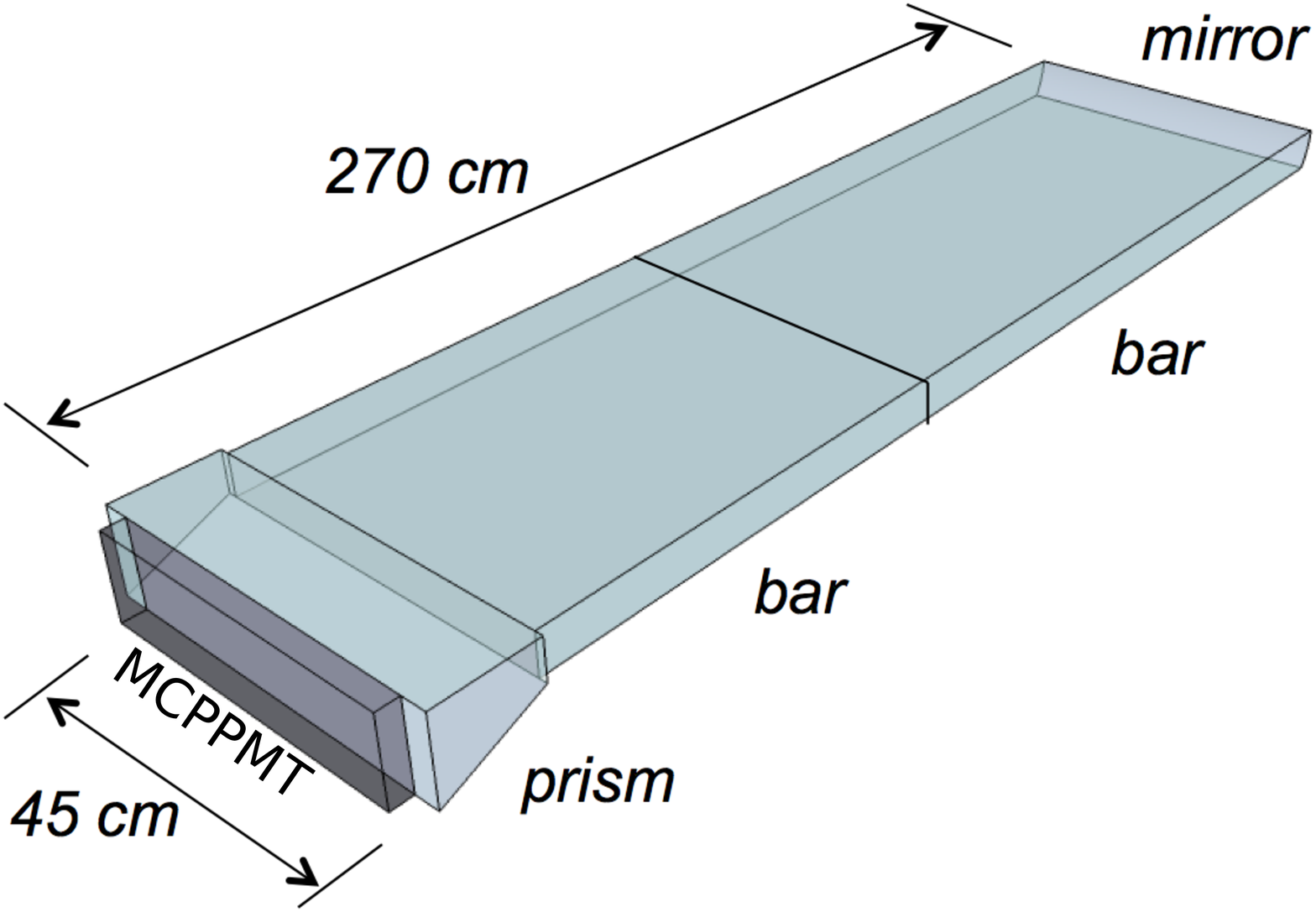}
\includegraphics[width=6cm]{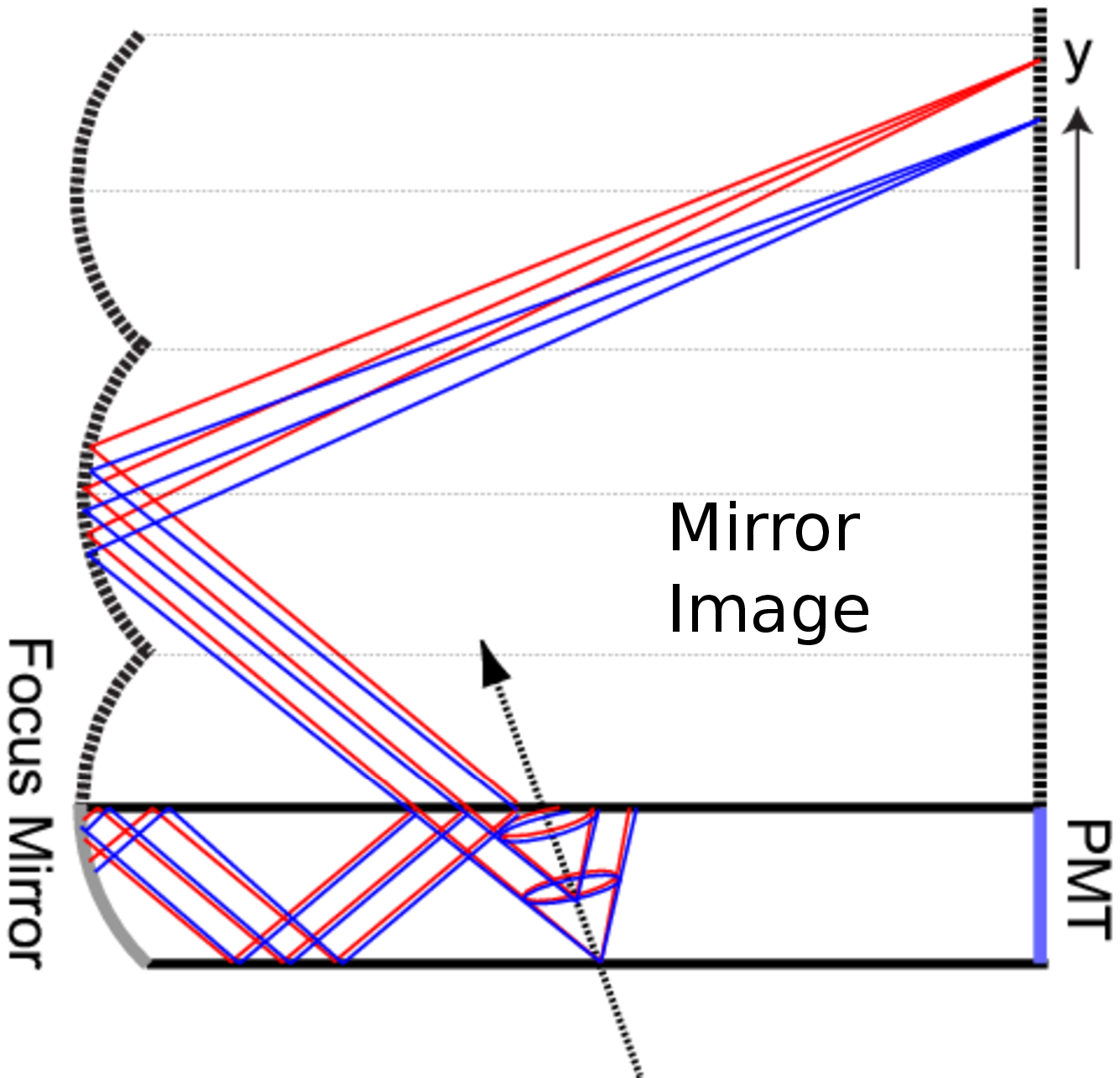}
\caption{A schematic diagram of the quartz radiator of the TOP counter is shown on the left. The need of the mirror in the counter is explained on the right. Parallel (non-parallel) rays are focused at the same (different) points of the detection plane by the mirror.}
\label{fig:module}
\end{figure}

A TOP counter module primarily consists of a quartz radiator bar, micro-channel plate photomultipliers (MCPPMTs) and a front-end readout. 
The Cherenkov photons emitted from a charged particle in the quartz bar are internally reflected toward the detection plane. 
The obtained Cherenkov ring image is characterized by its time of arrival and impact position in the detector plane~\cite{ratcliff}.
Two quartz bars each having dimension $(1250 \times 450 \times 20)~\rm mm^{3}$, a mirror with dimension $(100 \times 450 \times 20)~\rm mm^{3}$, and a small expansion prism of dimension $(100 \times 456 \times 20-51)~\rm mm^{3}$ are epoxied with an optical glue to form the quartz radiator, as shown in the Figure~\ref{fig:module} (left). 
The radiator is enclosed in a box made of aluminum honeycomb panels and is supported by PEEK polymer buttons to leave an air gap between the radiator and the panels.  
The mirror has a radius of curvature of about 6500 mm. As such, it focuses parallel rays in the Cherenkov cone to a single point and thus removes the effect of the bar thickness (see Figure~\ref{fig:module} (right)). 
Non-parallel rays are focused to different points of the detector plane, which gives a better charged particle separation and also allows to correct for chromatic dispersion. 
Optical components of the radiator should be of high quality to prevent any photon loss during propagation or reflections. To ensure this, several quality acceptance requirements are needed~\cite{boqun}. 

Cherenkov photons, emitted in the radiator due to passage of a charged particle, go through total internal reflections and are finally register at the expansion volume end by an array of MCPPMTs. Different path lengths are traversed by the photons depending on $\theta_{c}$, which results in a different time of propagation. The arrival time, including the time of flight of the charged particle and the time of propagation of emitted photons, and position in the detector plane of each detected Cherenkov photon are used to compute a likelihood for a given particle mass hypothesis.
 
A square-shaped MCPPMT with an effective area $(23 \times 23) ~\rm mm^{2}$ has been developed in collaboration with Hamamatsu Photonics KK for the TOP counter~\cite{inami}\cite{akatsu}. 
It has a multi-alkali photocathode whose average quantum efficiency is about 28\% for wavelengths around 380 nm. In the MCPPMT, there are two multi-channel plates (MCP) of thickness $400~\rm\mu m$ with a bore size diameter of $10~\rm\mu m$.
The bias angle of the pore is $13^{\circ}$. 
The readout from each MCPPMT anode is divided into $4 \times 4$ channels of area $(5.3 \times 5.3) ~\rm mm^{2}$. 
Each module contains two rows of 16 MCPPMTs, giving in total 32 MCPPMTs or 512 anode channels per module. 
The photo-electron emitted by the photo-cathode is multiplied in the MCP pores and then detected at one of the anode channel. 
The transit time spread is about 35~ps. 
By applying the nominal operating bias, a gain of $10^{6}$ is achieved.

The MCPPMTs are physically mounted on the front-end electronics modules. 
The main components of the front-end electronics are: front board to host the MCPPMT array; high voltage board to provide high voltages to the MCPPMTs; Application-Specific Integrated Circuits (ASICs); and Standard Control, Read-Out, and Data (SCROD) board. 
The ASICs consist of switched-capacitor array memory for the fast sampling and storage of time-sampled analog waveforms. 
The SCROD is the main controller of the entire module and acts as an entry/exit command/data point to the unit~\cite{andrew:ieee}\cite{andrew:tipp2014}.

\begin{figure}[h]
\centering
\includegraphics[width=12cm]{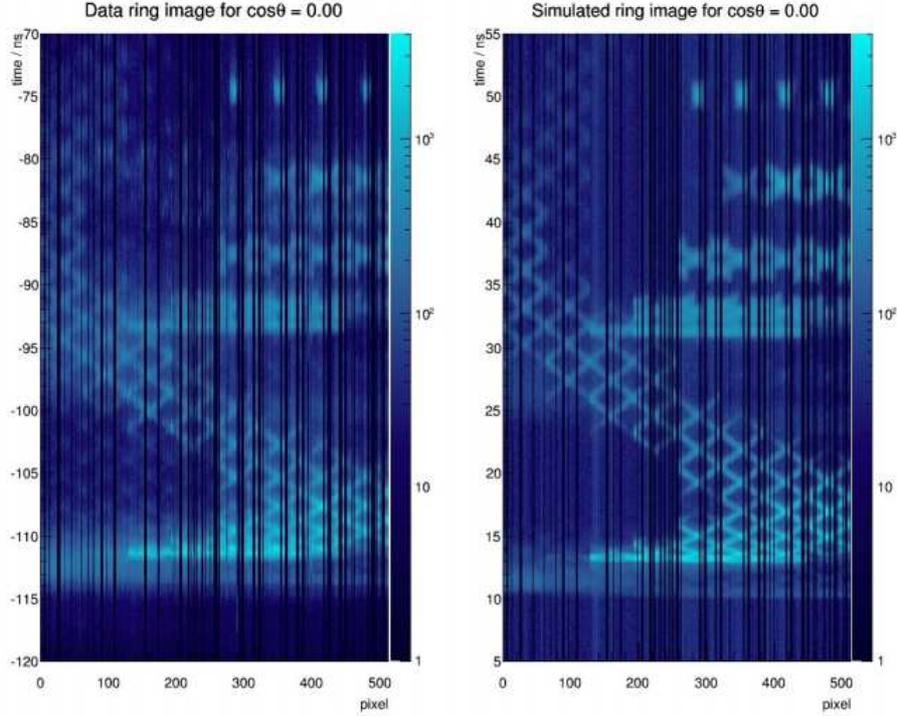}
\caption{Distributions of time (ns) {\it vs.} MCPPMT channels, in data from the beam test at SPring-8 (left) and simulations (right) for the case of beam hitting the TOP prototype at a normal incidence.}
\label{fig:beamtest}
\end{figure}

A small prototype of the TOP counter was tested at the $1.2~\rm GeV/c$ positron beam at LEPS (Laser Electron Photon beam line at SPring-8) in June 2013. 
For the beam test, the positron beam was tracked by the LEPS spectrometer and the beam timing was obtained from the accelerator RF. 
Data were taken with the beam hitting the prototype at normal incidence and also at some angle. 
In the latter case, the emitted Cherenkov photons will mostly be reflected from the mirror and then reach the detector plane. For both cases of beam incidence, results in data and simulations were found to be in agreement.
In a single event, the number of detected photons was 20-30. 
An integrated distribution in data and simulations over many events for the normal incidence case is shown in Figure~\ref{fig:beamtest}.

\section{ARICH counters}
\label{epid}

\begin{figure}[h]
\centering
\includegraphics[width=15cm]{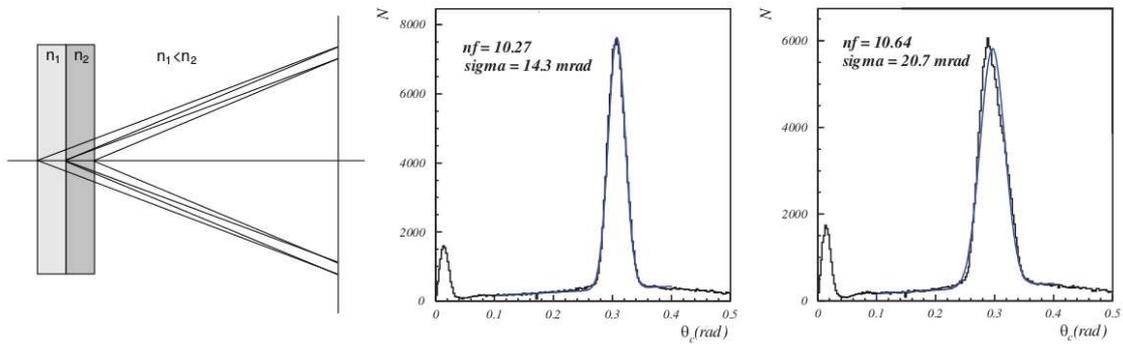}
\caption{Schematic of the focusing scheme with two aerogel layers of different refractive indices (left). Number of Cherenkov photons {\it vs.} $\theta_{c}$ for the focusing configuration (center) and for a 40~mm single homogeneous aerogel layer (right)~\cite{korpar}.}
\label{fig:FocConf}
\end{figure}

The ARICH counter is basically a proximity focusing RICH. Its main components are: aerogel tiles as a radiator, an array of position sensitive photon detectors, and a readout system~\cite{matsumoto}\cite{iijima}\cite{nishida}. 
The aerogel radiator consists of two layers with increasing refractive indices along the particle path so that the Cherenkov photons emitted by each layer overlap at the photo-detection plane, as shown in Figure~\ref{fig:FocConf} (left). 
The two aerogel tiles are of thickness 20~mm each and have refractive indices of 1.045 and 1.055 for upstream and downstream tiles, respectively. 
This arrangement of two layers with different refractive indices gives better performance than a single aerogel layer for the entire thickness (40~mm), as the latter arrangement leads to degradation of the single photon resolution because of an increase in emission point uncertainty. 
The single photon resolution ($\sigma_{\theta}\sim 14~\rm mrad $) for the dual radiator is significantly better than for the case of single radiator ($\sigma_{\theta}\sim 21~\rm mrad $) without affecting the number of photons detected (Figure~\ref{fig:FocConf}).

The photo-detector for the ARICH should be sensitive to single photon detection, able to provide position information, be immune to the 1.5~T magnetic field perpendicular to the photon detection plane, and be tolerant of the high radiation environment. 
A Hybrid Avalance Photo-Detector (HAPD) was developed in a joint effort with Hamamatsu. The HAPD consists of a bialkali photocathode, a vacuum region with a high electric field, and four avalanche photo-diode (APD) chips that are further pixelated in $6\times 6$ pixels giving in total 144 pixels of size ($4.9 \times 4.9 ~\rm mm^{2}$).
An incident photon is coverted into a photo-electron by the photocathode, which has a peak quantum efficiency of about 28\% at 400~nm.
The photo-electron subsequently accelerates towards the pixelated APDs via a high electric field and gets multiplied due to bombardment gain by a factor of about 1800. 
An additional gain of about 40 is achieved from the avalanche process of the APDs, resulting an overall gain of 70,000.
The ARICH covers $3.5~\rm m^{2}$ in the forward endcap of the Belle II detector; the aerogel radiator plane and the photo-detection plane are separated by a distance of 200~mm.
The radiator plane consists of 124 pairs (dual-layer, in total 248) of aerogel tiles while the photo-detection plane contains 420 HAPDs. At the outermost edge of these two planes, planar mirrors (in total, 18) are placed to redirect the outside-going photons towards the detection plane.
Dedicated high gain and low noise electronics were developed for the readout.
To each HAPD a front-end board with four ASICs and a field programmable gate array is attached.
The digitized hit information is collected by a merger board from front-end boards, and then communicated to further stages of the data acquisition system~\cite{nishida:readout}.

\begin{figure}[h]
\centering
\includegraphics[width=8cm]{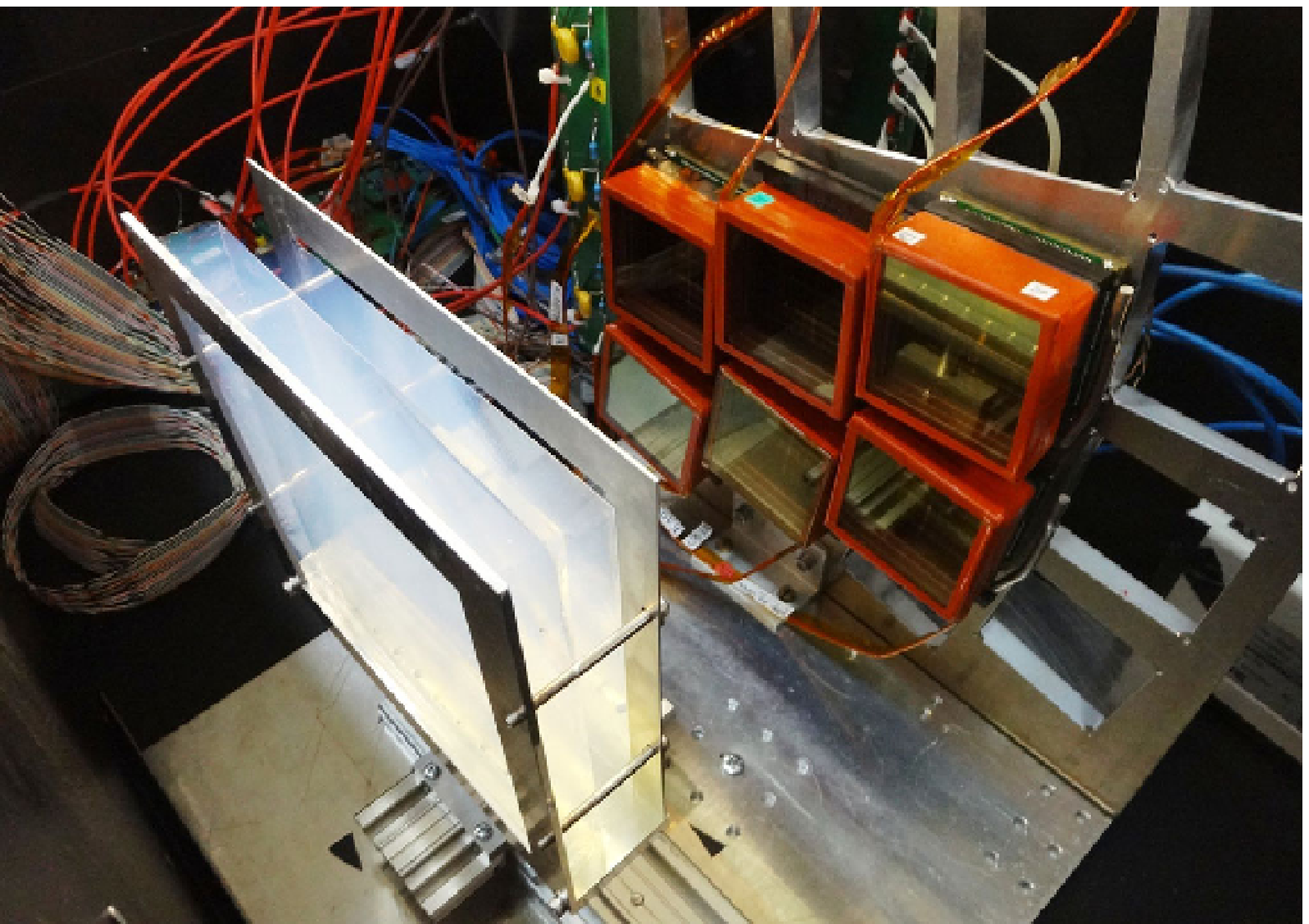}
\includegraphics[width=6.5cm]{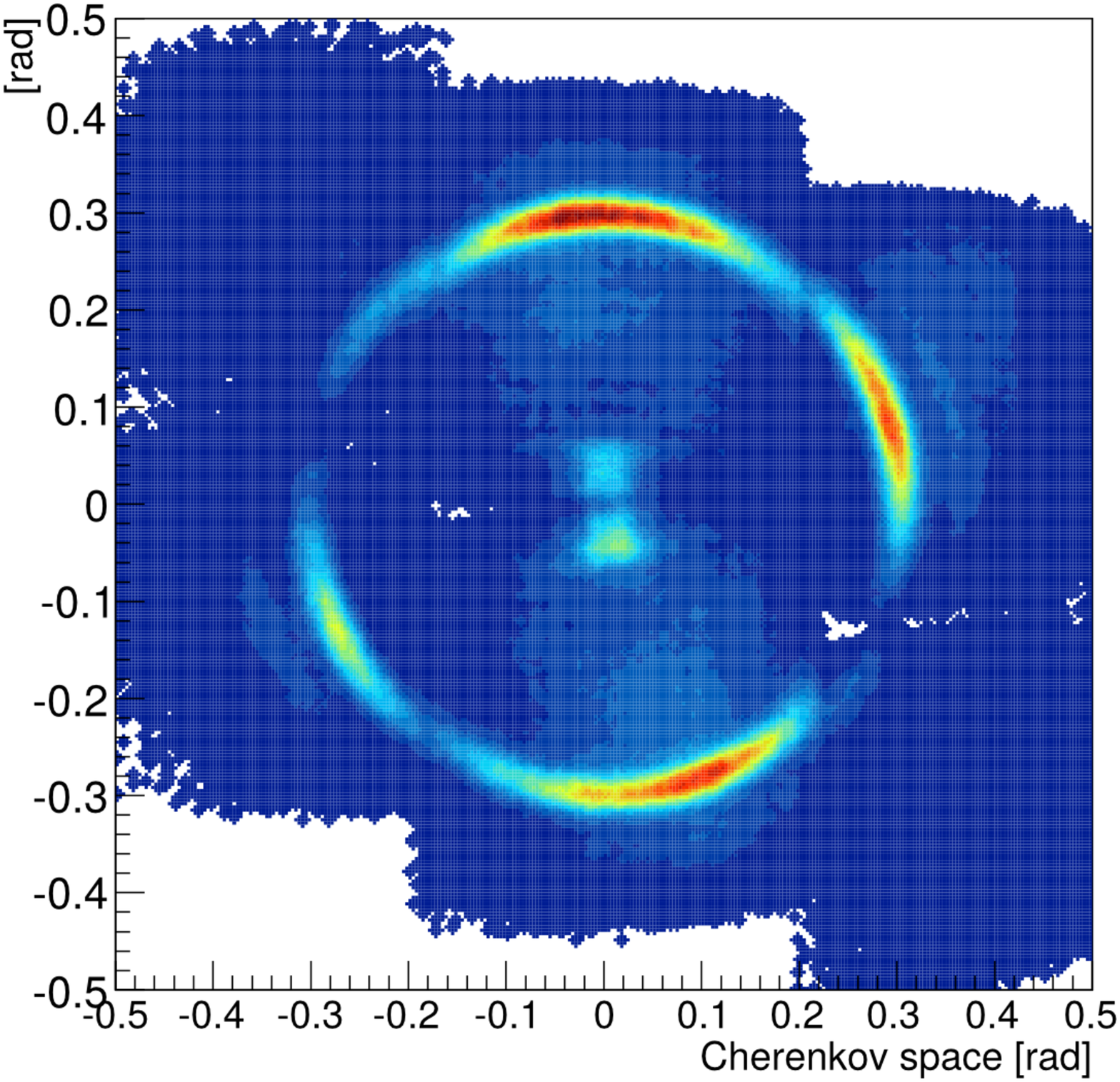}
\caption{Photograph of the prototype of ARICH for the DESY beam-test(left). Accumulated photon hit distribution with respect to the track position (right).}
\label{fig:arich_bt}
\end{figure}

A prototype ARICH was tested at the DESY test beam in May 2013. 
A photograph of the setup is shown in Figure~\ref{fig:arich_bt} (left). 
The accumulated hit distribution with respect to the track position is shown in  Figure~\ref{fig:arich_bt} (right), in which a Cherenkov ring is clearly visible.
The Cherenkov angle resolution is found to be 15.8~mrad, and on average 9 photons per track are detected. 

\section{Summary and Status}
\label{summary}
The TOP and ARICH detectors will provide PID information in the barrel and forward endcap region, respectively, of the Belle II detector. 
The working principle of both the detectors is based on imaging the Cherenkov ring created by the passage of charged particles. 
The TOP utilizes the impact position and time of arrival of the Cherenkov photons at the detection plane after total internal reflections in the quartz radiator bar to obtain PID information, whereas the ARICH is a proximity focusing RICH detector.
Prototypes of both the detectors demonstrated their expected performance in test-beams. 
All 16 TOP counters have been successfully installed in the Belle II detector.  
Installation of the HAPDs and the aerogel tiles has already started and will be finished by the Summer of 2016.

Detailed simulations have been performed with the Belle II software framework, based on which we expect an excellent charged kaon efficiency ($>90\%$) with a very small ($<10\%$) charged pion misidentification probability.

\end{document}